\documentclass  {article}
\title{DNA and the double helix: statistical equilibrium and Neumann's principle}
\author{Michael Caola\\6 Normanton Rd., Bristol, BS8 2TY, U.K.\\caola@blueyonder.co.uk}
\begin{document}

\maketitle
\begin{abstract}

Neumann's principle (that the symmetry of a crystal measurement cannot be lower than that of its point-group) is a corner- 
stone of crystallography: were it false, then the technique of x-ray diffraction (double-helix, DNA) might well not exist. The literature variously regards its truth as obvious, intuitive, axiomatic or even impossible [10], without further analysis or proof. After identifying and correcting a false lead/start, we give a plausible proof of Neumann's principle, using group theory and quantum statistical mechanics.

\end{abstract}
\section{Preamble}
Neumann's principle, that the symmetry of any measurable property of a crystal cannot be lower than that of its point-group, is a cornerstone of crystallography [7-10]. Its truth is often assumed obvious or at least plausible, with no further analysis or proof: indeed [10] states that no proof is possible. Against this background we give some analysis and a plausible proof, using group theory and the fundamental postulate of statistical mechanics
\section{Introduction}
Consider a finite macroscopic body whose electronic Hamiltonian $H$ has the symmetry of point group $G$. Its eigenfunctions are $|s\alpha i\rangle$, where $\Gamma^{(\alpha)}$ is an irreducible representation of $G$, $i=1,2,$ .. $d_{\alpha}$ labels the energy degeneracy and $s$ distinguishes levels with the same $\alpha$: $H|s\alpha i\rangle=E_{s\alpha}|s\alpha i\rangle$ [1]. The most general quantum state $|\psi\rangle$ of the body is a superposition of its eigenfunctions $|s\alpha i\rangle$:
\begin{equation}
|\psi\rangle=\sum_{s\alpha i}a_{s\alpha i}|s\alpha i\rangle
\end{equation}
The body's symmetry is that of the density of electronic matter in it, $\rho=\psi\psi*$, where $\psi(r)=\langle r|\psi\rangle$ is its wave function in cooordinate space $r$ [2]. But the symmetry of $\psi_{s\alpha i}(r)=\langle r|s\alpha i\rangle$ is generally lower than $G$: for instance, we accept both that a hydrogen atom has spherical symmetry and that the electron density of a $p$-state is less than spherical. The same applies to all point groups $G$: the symmetry of a degenerate state $|s\alpha i\rangle$ ({\it i.e.} $d_{\alpha}>1$) is generally less than $G$ [1]. 
\par
Thus the symmetry of a general state $|\psi\rangle$, which is an arbitrary superposition of states, many of which have lower symmetry that $G$, is lower than its Hamiltonian $H$. This last sentence is worrying since we in fact determine the symmetry of $H$ by experimentally observing the symmetry of the body (e.g. by X-ray diffraction from the electron density $\rho$); this means that observed states of the body have the symmetry of $H$, whereas the general state $|\psi\rangle$ has lower symmetry. The actual state $|\psi_{obs}\rangle$ realised by the body must then be a special case of the the general state $|\psi\rangle$. How does this come about? We shall show that it is because the body is in statistical equilibrium, which is the effective condidion of most observations.

\section{Analysis}
\par
We consider that the body is in statistical equilibrium and accept the fundamental postulate of equal a priori probabiliities and random phases in statistical mechanics [3]. This mean here that $a_{s\alpha i}=a_{s\alpha}\exp i\phi_{s\alpha i}$ where $\phi_{s\alpha i}$ is the random phase and $\omega_{s\alpha}=|a_{s\alpha}|^2$ is the probability of finding the body in state $|s\alpha i\rangle$. Now the observed density is the ensemble average value of $\rho$:
\begin{equation}
\rho(r)=\sum_{s\alpha i s'\alpha' i'}a_{s\alpha i}a_{s'\alpha' i'}^{*}\langle s'\alpha' i'|r\rangle\langle r|s\alpha i\rangle,
\end{equation}
so its ensemble average is

\begin{equation}
\overline{\rho(r)}=\sum_{s\alpha i s'\alpha' i'}\overline{a_{s\alpha}a_{s'\alpha'}^* }\overline{\exp i[\phi_{s\alpha i}-\phi_{s'\alpha' i'}]}\psi_{s\alpha i}(r)\psi_{s'\alpha' i'}(r)^*.
\end{equation}
Since random phases mean that $\overline{\exp i[\phi_{s\alpha i}- \phi_{s'\alpha' i'}]}=\delta_{ss'}\delta_{\alpha\alpha'}\delta_{ii'}$, then
\begin{equation}
\overline{\rho(r)}=\sum_{s\alpha}\omega_{s\alpha}\sum_{i}\psi_{s\alpha i}\psi_{s\alpha i}^*
\end{equation}
But there is a theorem in group theory [1] which states that
\begin{equation}
\rho_{s\alpha}(r)=\sum_{i}\psi_{s\alpha i}\psi_{s\alpha i}^*
\end{equation}
transforms as $\Gamma^{(1)}$, the unit (identical) irreducible representation of $G$. This alone of the $\Gamma^{(\alpha)}$is invariant under all the transformations of $G$, and thus has the same high symmetry as the Hamiltonian $H$. Thus $\overline{\rho(r)}=\sum_{s\alpha}\omega_{s\alpha}\rho_{s\alpha}(r)$ has the same symmetry as $H$, which is what we wanted to show.

\section{External perturbations and time averages}
We have implicity assumed until now that the Hamiltonian $H$ of the body is time-independent. This is not so and to this extent our analysis so far is misleading (though conventional). The Hamiltonian is in fact $H+V(t)$, where $V$ it the interaction of the body with its surroundings --- the rest of the world [2,4]. It is this time-dependent perurbation $V$ which establishes equilibrium, and the corresponding state is time-dependent:
\begin{equation}
|\psi(t)\rangle=\sum_{s\alpha i}a_{s\alpha i}(t)|s\alpha i\rangle
\end{equation}
Any observation of the the density $\rho$ takes a finite time $\tau$ and we thus observe a time average of the instantaneous density $\rho(r,t)$:
\begin{equation}
\overline{\rho}(r,t)=\frac{1}{\tau}\int_{t}^{t+\tau}\rho(r,t')dt'
\end{equation}
We accept the common assumption (ergodic hypothesis) that time and ensemble averages are equivalent in statistical equilibrium [3]. The fundamental postulate for our time regime is then $a_{s\alpha i}(t)=a_{s\alpha}\exp i\phi_{s\alpha i}(t)$, and this could be produced by a $V$ that is random and stationary. The randomness of $V$ gives the random phases $\phi_{s\alpha i}$ and its stationarity gives the $a_{s\alpha}$ independent of time for observations lasting $\tau$ [5,6]. Thus (7) becomes
\begin{equation}
\overline{\rho}(r,t)=\overline{\rho}(r)=\sum_{s\alpha i s'\alpha' i'}a_{s\alpha}a_{s'\alpha'}\psi_{s\alpha i}(r)\psi_{s'\alpha' i'}(r)^*\frac{1}{\tau}\int_{t}^{t+\tau}\exp i[\phi_{s\alpha i}(t')-\phi_{s'\alpha' i'}(t')]dt'
\end{equation}
Since the phase is random
\begin{equation}
\frac{1}{\tau}\int_{t}^{t+\tau}\exp i[\phi_{s\alpha i}(t')-\phi_{s'\alpha' i'}(t')]dt'=\delta_{ss'}\delta_{\alpha\alpha'}\delta_{ii'}
\end{equation}
and the density $\overline{\rho}(r)=\sum_{s\alpha}\omega_{s\alpha}\rho_{s\alpha}(r)$, as before.

\section{Neumann's principle}
This states that the symmetry of any measurable physical property of a crystal cannot be lower than that of the crystal point group [7-10]. The above discussion for a general macroscopic body may thus be said to be a justification of Neumann's principle. We remark that the statistical equilibrium is usually, though not necessarilly, thermal: probability $\omega_{s\alpha}=e^{-E_{s\alpha}/kT}/\sum_{s\alpha}e^{-E_{s\alpha}/kT}
$.

\section{Discussion}
We showed in $\S$2 that a macroscopic body in statistical equilibrium has the symmetry of its Hamiltonian $H$; this followed on acceptance of the postulate of equal a priori probabilities and random phases. This postulate is commonly introduced without suggesting a physical mechanism to justify it, a justification being that that it gives the right answers [3]. One further has the impression that the posulate is supposed to hold for a truly isolated macroscopic body, {\it i.e.} one with time-independent Hamiltonian $H$. Extrapolation from the ensemble average of a large number of hypothetical, 'similar', isolated bodies to the actual behaviour of one observed body then requires an ergodic hypothesis identifying ensemble- and time-averages.
\\\\
But there is a school of thought which say that no macroscopic body can be truly isolated from its surroundings and that the interaction $V(t)$ is the mechanism responsible for statistical equilibrium [2,4]: the separation between energy levels of a macroscopic body is very small compared to the perturbation $V$ caused by the rest of the world, and it is this time-dependent perturbation which causes the transitions responsible for the establishment of statistical equilibrium. They say that it is natural to assume that $V$ is random and stationary [2,4,5,6].
\\\\
We do not wish to examine the bases of statistical mechanics, in particular ergodic hypotheses. The proof in {\S}2 is not modified by {\S}3, which offers a mechanism making the fundamental postulate plausible. Sections {\S}2 and {\S}3 show that a macroscopic body has the symmetry of its Hamiltonian regardless of whether one believes it to be isolated and governed by the fundamental postulate or to be necessarily in interaction with the rest of the world.

\section*{References}
[1] Tinkham, M., 'Group theory and qunatum mechanics', Mcgraw Hill, New York (1964)\\\\
$[2]$ Landau, L. D. and Lifchitz, E. M., 'Statistical physics', Pergamon Press, London (1958)\\\\
$[3]$ Tolman, R. C., 'The principles of statistical mechanics', Oxford University Press, Oxford (1938)\\\\
$[4]$ Blatt, J. M., Prog Theor. Phys., {\bf 22} 745 (1959)\\\\
$[5]$ Abragam, A., 'The principles of nuclear magnetism', Oxford University Press, Oxford (1961)\\\\
$[6]$ Slichter, C. P., 'Principles of magnetic resonance', Harper and Row, New York (1963)\\\\
$[7]$ Bhagavantam, S., 'Crystal symmetry and physical properties', Academic Press, New York (1966)\\\\
$[8]$ Smith, C. S., Solid State Physics {\bf 6}, 175 (1958)\\\\
$[9]$ Birss, R. R., Rep. Prog. Phys., 312 (1963)\\\\
$[10]$ Burns, G., 'Solid state physics', Academic Press, San Diego (1985)

 \end{document}